\documentclass[10pt,english,aps,prd,preprintnumbers,superscriptaddress,floatfix,showpacs,nofootinbib]{revtex4}
\usepackage{graphicx}
\usepackage{hyperref}
\usepackage{amsmath}
\usepackage{amssymb}

\newcommand{\be}{\begin{eqnarray}}
\newcommand{\ee}{\end{eqnarray}}
\newcommand{\non}{\nonumber\\}

\newcommand{\ave}[1]{\left\langle #1 \right\rangle}

\newcommand{\xp}{x}
\newcommand{\xm}{\bar{x}}
\newcommand{\Np}{N}
\newcommand{\Nm}{\bar{N}}
\newcommand{\np}{n}
\newcommand{\nm}{\bar{n}}
\newcommand{\ep}{\epsilon}
\newcommand{\emm}{\bar{\epsilon}}

\usepackage{color}


\begin{document}

\title{Local Efficiency Corrections to Higher Order Cumulants} 

\author{
A. Bzdak
}
\email[E-Mail: ]{abzdak@quark.phy.bnl.gov}
\affiliation{RIKEN BNL Research Center, Brookhaven National Laboratory, 
Upton, NY 11973, USA
}
\affiliation{AGH University of Science and Technology, 30-059 Krak\' ow, Poland
}

\author{
V. Koch
}
\email[E-Mail: ]{vkoch@lbl.gov}
\affiliation{
Nuclear Science Division,
Lawrence Berkeley National Laboratory, 1 Cyclotron Road,
Berkeley, CA 94720
}
\date{\today}

\begin{abstract}
In this brief note we derive and present the formulas necessary to
correct measurements of cumulants for detection efficiency. 
In particular we consider the case where the
efficiency may depend on the phase-space, such as transverse momentum,
rapidity etc. 
\end{abstract}

\maketitle

\section{Introduction}
The study of the phase structure of the strong interaction has
been a central topic in strong interaction physics for many
years. Lattice quantum chromodynamics (QCD) calculations have meanwhile established that the
transition at vanishing net-baryon density is an analytic cross-over
transition \cite{Aoki:2006we}. The situation at finite net-baryon
density, however, is not yet settled. Indeed many effective models for the
strong interaction (see Ref. \cite{Stephanov:2007fk} for an overview)
predict a first order phase-coexistence line which ends in a critical
point, the location of which is rather model dependent. 

In order to explore the QCD phase diagram in experiment, a
beam energy scan has been carried out at the Relativistic Heavy Ion
Collider (RHIC). By varying the beam energy the temperature and the
net-baryon density of the system created in a heavy-ion reaction can
be changed, with collisions at lower energies leading to system at
higher net-baryon density. Among many observables the cumulants of
the proton number distribution has received particular attention,
since they can be considered to be a measure for net-baryon number
fluctuations \cite{Hatta:2003wn}, which in turn are sensitive to the presence
of any structure in the QCD phase diagram at finite net-baryon
density \cite{Skokov:2010uh, Stephanov:2011pb}. The analysis and interpretation of
these cumulants, however, need to be carried out with some
care. Effects such as baryon number conservation \cite{Bzdak:2012an},
the non-observation of neutrons \cite{Kitazawa:2012at}, and
efficiency corrections \cite{Bzdak:2012ab,Garg:2012nf} 
need to be taken into account before any
conclusions on possible phase changes can be drawn from the data. 

The purpose of this short note is to revisit the corrections due to
finite detection efficiency, $\epsilon$, on particle number cumulants,
as discussed in \cite{Bzdak:2012ab}, and extend the formalism to
include a possible dependence of the efficiency on the phase space,
such as transverse momentum, rapidity or azimuthal angle. 

Before we start let us recall why the detection efficiency $\epsilon$
leads to corrections of any fluctuation observable, such as particle
number cumulants of order $n \geq 2$. Suppose in each event we have
exactly $N$ particles, i.e., the variance of the number of {\em
  produced} particles vanishes. 
Given a detection efficiency $\epsilon$, the mean number of {\em
  observed} particles, $n$, is then $\ave{n}= \epsilon N$. 
However, this does {\em not} imply that in
each event $i$ the number of observed particles is $n_i =\epsilon
N$. Instead, $n_i$ fluctuates around the mean $\ave{n}$ so that the 
distribution of observed particles has a finite variance. More
formally, assume that the number of produced particles $N$
is distributed according to $P(N)$. Then the number of
observed particles follows a distribution $p(n)=\sum_N w(n|N) P(N)$
where $w(n|N)$ denotes the probability to observe $n$ particles given
$N$ produced particles. Clearly, in general, the cumulants of $P(N)$
and $p(n)$ are different. Typically $w(n|N)$ is modeled by a binomial
distribution, and a recent analysis of net-proton cumulants by the
STAR collaboration employed this approach \cite{Adamczyk:2013dal}.

Efficiency corrections to variances of the net-charge and net-baryon
distributions have been studied in
\cite{Koch:2001zn,Pruneau:2002yf,Bower:2001fq} and have been extended
to higher order cumulants in 
\cite{Bzdak:2012ab,Garg:2012nf}.  
In all these studies a constant efficiency $\epsilon$ has been
assumed. In reality, however, the efficiency may very well depend
on the kinematics of the particles, such as their transverse momentum,
rapidity and azimuthal angle. This will likely result in further corrections
especially for higher order cumulants. 
It is the purpose of this short note to derive the necessary formulae
that relate the moments and cumulants of the distribution of {\em produced}
particles to those of the {\em observed} particles for the situation
of phase-space dependent efficiencies. As in Ref. \cite{Bzdak:2012ab} we will
assume that the (phase-space dependent) detection probabilities may be
modeled by binomial distributions. 

\section{Local Efficiency Corrections}
\label{sec:effic-corr}
Let us start by introducing some notation. In general we will use upper case
characters to refer to the {\em produced} particles and lower case for
the {\em observed}
particles. Since we are 
interested in cumulants and moments of the net-baryon number, net-proton number,
or net-charge, we will have two types of particles such as
baryon/anti-baryon, proton/anti-proton, positive/negative
charge. Thus we will denote the number of produced particles and
``anti''-particles\footnote{We use the term anti-particles quite
  generally. For example, in case of net-charge, ``particles'' refer
  to positively charged particles and ``anti-particles'' refer to
  negatively charged particles.} by  $\Np$ and $\Nm$, respectively, and the observed
ones by $\np$ and $\nm$. Next, we assume that the phase-space will be partitioned
into bins of potentially varying size. The location of these bins in
rapidity, transverse momentum and azimuthal angle, $[y,p_\perp,\phi]$, will
be denoted by $\xp$ and $\xm$ for particles and anti-particles, respectively.   
Thus $\Np(\xp)$ denotes the number of {\em produced}
particles in the phase-space bin located at $\xp$, while $\np(\xp)$ is
the number of {\em observed} particles at $\xp$, etc.
The event averaged number of produced and observed
particles in the phase-space bin located at $\xp$ are then given by
$\ave{\Np(\xp)}$ and $\ave{\np(\xp)}$, respectively.  Similarly
$\ave{\Np(\xp)\Nm(\xm)}$ is the event average of the product of the
number of particles at $\xp$ and anti-particles at $\xm$. In
order to obtain the event average over {\em all} the particles in the
considered phase-space we need to sum over all phase-space bins
\begin{eqnarray}
  \label{eq:1}
  \ave{\Np}&=&\sum_{\xp}\ave{\Np(\xp)}\\
  \ave{\Np\Nm}&=&\sum_{\xp}\sum_{\xm}\ave{\Np(\xp)\Nm(\xm)}\\
  \label{eq:1c}
  \ave{\Np^{i}\Nm^{k}}&=&\sum_{\xp_{1},\ldots,\xp_{i}}
  \sum_{\xm_{1},\ldots,\xm_{k}}
  \ave{\Np(\xp_{1})\ldots\Np(\xp_{i})\Nm(\xm_{1})\ldots \Nm(\xm_{k})}
\end{eqnarray}
and similarly for the observed particles.

Next we introduce the probability $w(n(x)|N(x);\epsilon(x))$ to
observe $n(x)$ particles in the phase-space bin at $x$ given $N(x)$
produced particles and a phase-space dependent detection efficiency of
$\epsilon(x)$. The efficiencies for anti-particles are correspondingly 
$\emm(\xm)$. As already discussed we will model $w$ as a binomial
distribution, where the binomial probability is given by the efficiency
$\epsilon(x)$:
\begin{eqnarray}
  \label{eq:2}
  w(n(x)|N(x);\epsilon(x))=\frac{N(x)!}{n(x)!\left(N(x)-n(x)\right)!}
  \,\epsilon(x)^{n(x)}\left(1 - \epsilon(x)\right)^{N(x)-n(x)}.
\end{eqnarray}

Given $w(n(x)|N(x);\epsilon(x))$ we can relate the probability
%$p\left(\np(\xp_{1}),\ldots,\np(\xp_{n})\right)$ 
to {\em observe} a
given number of particles $n(x_{i})$ at the various phase space points
$x_{i}$ to the probability for produced particles at these
points, 
%$P\left(\Np(\xp_{1}),\ldots,\Np(\xp_{n})\right)$,  
\begin{eqnarray}
  \label{eq:3}
\lefteqn{p\left(\np(\xp_{1}),\ldots,\np(\xp_{n});\,\nm(\xm_{1}),\ldots,\nm(\xm_{m})\right) =  
 \sum_{\Np(\xp_{1})=\np(\xp_{1})}^{\infty}
%\ldots \sum_{\Np(\xp_{m})=\np(\xp_{n})}\ldots \sum_{\Nm(\xm_{1})=\nm(\xm_{1})}
\ldots \sum_{\Nm(\xm_{m})=\nm(\xm_{m})}^{\infty}
} \non &&
w\left(\np(\xp_{1})|\Np(\xp_{1});\ep(\xp_{1})\right) \ldots
w\left(\nm(\xm_{m})|\Nm(\xm_{m});\emm(\xm_{m})\right) \,
P\left(\Np(\xp_{1}),\ldots,\Np(\xp_{n});\,
\Nm(\xm_{1}),\ldots,\Nm(\xm_{m})\right).
\end{eqnarray}
Although the above expression looks rather involved, the relations between
the various moments of the observed and produced particle distributions are straightforward to write down. The reason is that the binomial
distributions  for the various bins in phase-space and between particles and
anti-particles are independent from each other. For the
lowest moments we get
\begin{eqnarray}
  \label{eq:4}
  \ave{\np(\xp)} &=& \ep(\xp) \ave{\Np(\xp)} \\
  \label{eq:4b}
  \ave{\np(\xp)\nm(\xm)} &=& \ep(\xp)\emm(\xm) \ave{\Np(\xp)\Nm(\xm)}\\
  \label{eq:4c}
  \ave{\np(\xp_{1})\np(\xp_{2})} &=& \ep(\xp_{1})\ep(\xp_{2})
  \ave{\Np(\xp_{1})\Np(\xp_{2})} \,\, \xp_{1}\neq \xp_{2}\\
  \label{eq:4d}
  \ave{\np(\xp)(\np(\xp)-1)} &=& \ep(\xp)^{2} \ave{\Np(\xp)(\Np(\xp)-1)}.
\end{eqnarray}
The last two equations, Eqs.~(\ref{eq:4c}),(\ref{eq:4d}), can be conveniently written as
\begin{eqnarray}
  \label{eq:4e}
  \ave{\np(\xp_{1})(\np(\xp_{2}) - \delta_{\xp_{1},\xp_{2}} )} &=& \ep(\xp_{1})\ep(\xp_{2})
  \ave{\Np(\xp_{1})(\Np(\xp_{2}) - \delta_{\xp_{1},\xp_{2}} )}
\end{eqnarray}
where $\delta_{\xp_{1},\xp_{2}}=1$ if $\xp_{1}=\xp_{2}$ and zero otherwise.  
In order to proceed and to arrive at a general relation between the
cumulants of the observed and produced particle distributions we
follow the same strategy as in our previous work \cite{Bzdak:2012ab}. 
There we expressed
the cumulants in terms of factorial moments and used a general
relation between the factorial moments of the distribution of observed
and produced particles. The factorial moments are given by
\begin{eqnarray}
  \label{eq:5}
  F_{i,k}\equiv \ave{\frac{\Np!}{(\Np-i)!}\frac{\Nm!}{(\Nm-k)!}}\\
  f_{i,k}\equiv \ave{\frac{\np!}{(\np-i)!}\frac{\nm!}{(\nm-k)!}}
\end{eqnarray}
for the distribution of produced and observed particles,
respectively. For constant, phase-space independent, efficiency
correction \cite{Bzdak:2012ab} 
\begin{eqnarray}
  \label{eq:6}
  f_{i,k}=\ep^{i}\emm^{k} F_{i,k}.
\end{eqnarray}
In order to allow for phase space depended efficiency
corrections we introduce the ``local'' factorial moments
\begin{eqnarray}
  \label{eq:7}
  A_{i,k}\left(\xp_{1},\ldots,\xp_{i};\xm_{1},\ldots,\xm_{k}\right)
  &=& \left<\Np(\xp_{1}) [\Np(\xp_{2})-\delta_{\xp_{1},\xp_{2}}]\ldots
    [\Np(\xp_{i})- \delta_{\xp_{1},\xp_{i}} - \ldots -   
    \delta_{{\xp_{i-1},\xp_{i}}}]\right. \non
    &&\left. \,\, \Nm(\xm_{1}) [\Nm(\xm_{2})-\delta_{\xm_{1},\xm_{2}}]\ldots
    [\Nm(\xm_{k})- \delta_{\xm_{1},\xm_{k}} - \ldots -
    \delta_{{\xm_{k-1},\xm_{k}}}] \right> \\
  a_{i,k}\left(\xp_{1},\ldots,\xp_{i};\xm_{1},\ldots,\xm_{k}\right)
  &=& \left<\np(\xp_{1}) [\np(\xp_{2})-\delta_{\xp_{1},\xp_{2}}]\ldots
    [\np(\xp_{i})- \delta_{\xp_{1},\xp_{i}} - \ldots -
    \delta_{{\xp_{i-1},\xp_{i}}}]\right. \non
    &&\left. \,\, \nm(\xm_{1}) [\nm(\xm_{2})-\delta_{\xm_{1},\xm_{2}}]\ldots
    [\nm(\xm_{k})- \delta_{\xm_{1},\xm_{k}} - \ldots -
    \delta_{{\xm_{k-1},\xm_{k}}}] \right>.
\end{eqnarray}
Using Eq.~\eqref{eq:1c}, it is straightforward to show that the
``local'' factorial moments, $A_{i,k}$ and $a_{i,k}$, are related to the factorial
moments $F_{i,k}$ and $f_{i,k}$ by summation over the phase-space bins
\begin{eqnarray}
  \label{eq:12}
  F_{i,k}&=&\sum_{\xp_{1},\ldots,\xp_{i}}\sum_{\xm_{1},\ldots,\xm_{k}}
  A_{i,k}\left(\xp_{1},\ldots,\xp_{i};\xm_{1},\ldots,\xm_{k}\right)\\
  f_{i,k}&=&\sum_{\xp_{1},\ldots,\xp_{i}}\sum_{\xm_{1},\ldots,\xm_{k}}
  a_{i,k}\left(\xp_{1},\ldots,\xp_{i};\xm_{1},\ldots,\xm_{k}\right)
\end{eqnarray}
Analogous to Eq.~\eqref{eq:6} the local factorial moments of
the observed particle distribution are related to that of the produced
particles by  
\begin{eqnarray}
  \label{eq:8}
  a_{i,k}=\ep(\xp_{1})\ldots\ep(\xp_{i})\emm(\xm_{1})\ldots\emm(\xm_{k}) A_{i,k}.
\end{eqnarray}
This relation follows from Eq.~(\ref{eq:6}) and the fact, that the
binomial efficiency corrections for different phase-space bins are
independent from each other. Clearly our generalized relation,
Eq.~(\ref{eq:8}), gives the correct results for the second order
moments, Eqs.~(\ref{eq:4}-\ref{eq:4b}).

By virtue of Eqs.~\eqref{eq:12} and
\eqref{eq:8}, the factorial moments of the produced particle distribution
can be extracted from the measured local particle distribution via
\begin{eqnarray}
  \label{eq:final_result}
  F_{i,k}=\sum_{\xp_{1},\ldots,\xp_{i}}\sum_{\xm_{1},\ldots,\xm_{k}}
\frac{a_{i,k}\left(\xp_{1},\ldots,\xp_{i};\xm_{1},\ldots,\xm_{k}\right)}
{\ep(\xp_{1})\ldots\ep(\xp_{i})\emm(\xm_{1})\ldots\emm(\xm_{k})}.
\end{eqnarray}
For example for $F_{2,2}$ we obtain
\begin{eqnarray}
  \label{eq:13}
  F_{2,2}&=&\sum_{\xp_{1},\xp_{2},\xm_{1},\xm_{2}}
  \frac{\ave{\np(\xp_{1})[\np(\xp_{2})-\delta_{\xp_{1},\xp_{2}}]\, 
    \nm(\xm_{1})[\nm(\xm_{2})-\delta_{\xm_{1},\xm_{2}}]}}
  {\ep(\xp_{1})\ep(\xp_{2})\emm(\xm_{1})\emm(\xm_{2})}\non
  &=& \sum_{\xp_{1},\xp_{2},\xm_{1},\xm_{2}} 
  \frac{\ave{\np(\xp_{1}) \np(\xp_{2}) \nm(\xm_{1}) \nm(\xm_{2}) }}
  {\ep(\xp_{1})\ep(\xp_{2})\emm(\xm_{1})\emm(\xm_{2})}
  -\sum_{\xp_{1},\xp_{2},\xm_{1}} 
  \frac{\ave{\np(\xp_{1}) \np(\xp_{2}) \nm(\xm_{1})  }}
  {\ep(\xp_{1})\ep(\xp_{2})\emm(\xm_{1})^2}\non
  && - \sum_{\xp_{1},\xm_{1},\xm_{2}} 
  \frac{\ave{\np(\xp_{1})  \nm(\xm_{1}) \nm(\xm_{2}) }}
  {\ep(\xp_{1})^{2} \emm(\xm_{1})\emm(\xm_{2})}
  + \sum_{\xp_{1},\xm_{1}} 
  \frac{\ave{\np(\xp_{1})  \nm(\xm_{1}) }}
  {\ep(\xp_{1})^{2} \emm(\xm_{1})^{2}} .
\end{eqnarray}

The relation, Eq.~\eqref{eq:final_result}, between the factorial
moments of the distribution of 
produced particles, $F_{i,k}$, and the local factorial moments of the observed particles, $a_{i,k}$, is the main result of this paper. To
extract cumulants from the factorial moments $F_{i,k}$ is
straightforward, and it has been discussed in \cite{Bzdak:2012ab}, where the
relevant formulas for cumulants up to the sixth order are provided. 

Finally, it is worth noticing that the number of terms in Eq. (\ref{eq:final_result}) is $m^{i+k}$, 
where $m$ is the number of bins.\footnote{To clarify the notation all $x_i$ and $\xm_k$ in
Eq. (\ref{eq:final_result}) are summed from the first bin to the $m$-th bin.} For example, for the 
fourth order cumulant $i+k=4$ leading to $m^4$ terms.

\section{Discussion}
\label{sec:discussion}

It would be interesting to get an idea about the magnitude of the corrections due to the local efficiency corrections. A
precise determination is very difficult, since the correction will depend on the true multiplicity distribution and on the specific distribution of the particles in phase-space. All we can attempt here is a rough estimate using certain, simplifying, assumptions. In order to keep the formalism manageable let us consider the variance of the distribution of positively charged particles only.  The extension to net-charge distribution is straightforward. The variance $\sigma^2$ is given in terms of the produced particles $N$ by
\begin{eqnarray}
\sigma^2 \equiv \ave{N^2}-\ave{N}^2 =  \sum_{x_1,x_2}\left[\ave{N\left(x_1\right)\left(N\left(x_2\right)-\delta_{x_1,x_2}\right)}-\ave{N\left(x_1\right)}\ave{N\left(x_2\right)}\right]+\sum_{x}\ave{N\left(x_1\right)}
\label{eq:correct_variance}
\end{eqnarray}
Next, we introduce a correlation function $C\left(x_1,x_2\right)$ such that
\begin{equation}
\ave{N\left(x_1\right)\left(N\left(x_2\right)-\delta_{x_1,x_2}\right)}=\ave{N\left(x_1\right)}\ave{N\left(x_2\right)}\left(1+C\left(x_1,x_2\right)\right). 
\label{eq:Corr_Func}
\end{equation}
The correlation function $C\left(x_1,x_2\right)$ controls the correlations of the particles in phase-space. In its absence, $C=0$, particles are distributed according to a Poisson distribution in each bin, and there are no bin-to-bin correlations. In this case, as we shall see, there is no difference between local and global efficiency corrections. Given the correlation function  $C\left(x_1,x_2\right)$ the variance can be expressed as
\begin{eqnarray}
\sigma^2  & = & \sum_{x_1,x_2}\ave{N\left(x_1\right)}\ave{N\left(x_2\right)}C\left(x_1,x_2\right)+\ave N=\ave N+\delta
\end{eqnarray}
where $\delta$ denotes the deviation from Poisson behavior. 

In order to see the difference between local and global efficiency corrections, let us suppose we measure the variance but only correct for the global or rather mean efficiency, $\bar{\epsilon}$, which is given by
\begin{eqnarray}
\bar{\epsilon} =\frac{\sum_{x}\epsilon(x)\ave{N(x)}}{\sum_{x}\ave{N(x)}}=\frac{\sum_{x}\epsilon(x)\ave{N(x)}}{\ave N}
\label{eq:eps_bar}
\end{eqnarray}
Using the expression derived in \cite{Bzdak:2012ab} we would extract the following for the variance 
\begin{eqnarray}
\bar{\sigma}^2=\frac{1}{\bar{\epsilon}^{2}}\ave{n\left(n-1\right)}-\frac{1}{\bar{\epsilon}^2}\ave{n}^2+\frac{1}{\bar{\epsilon}}\ave{n} = \frac{1}{\bar{\epsilon}^{2}}\sum_{x_1,x_2}\left[\ave{n\left(x_1\right)\left(n\left(x_2\right)-\delta_{x_1,x_2}\right)}-\ave{n\left(x_1\right)}\ave{n\left(x_2\right)}\right]+\frac{1}{\bar{\epsilon}}\sum_{x_1}\ave{n\left(x_1\right)}. 
\label{eq:sig_bar}
 \end{eqnarray}
The difference between the true variance, $\sigma^2$, which we would recover by applying local efficiency corrections, and that extracted by correcting only for the average efficiency, $\bar{\sigma}^2$, will be a measure for the importance of local efficiency corrections. In order to proceed, we express observed local moments in Eq. (\ref{eq:sig_bar}) by the true moments following the relations derived above, Eqs. (\ref{eq:4}--\ref{eq:4d})
\begin{eqnarray}
\bar{\sigma}^2 = \ave{N}^2 \frac{\sum_{x_1,x_2}\epsilon\left(x_1\right)\epsilon\left(x_2\right)\ave{N\left(x_1\right)}\ave{N\left(x_2\right)}C\left(x_1,x_2\right)}{\left[ \sum_{x_1}\epsilon\left(x_1\right) \ave{N\left(x_1\right)} \right]^2}+\ave N=\bar{\delta}+\ave N,
\end{eqnarray}
where $\bar{\delta}$ denotes again the deviation from Poisson. We note, without correlations, $\delta=\bar{\delta}=0$ and there is no difference between local and global efficiency corrections.

In order to estimate the difference between $\delta$ and $\bar{\delta}$ we further assume that the true particles distribution, $\ave{N(x)}$, the efficiency, $\epsilon(x)$ and the correlation function, $C\left(x_1,x_2\right)$ are all given by Gaussians
\begin{eqnarray}
\ave{N\left(x\right)} & = & \frac{N_{\rm tot}}{\sqrt{2\pi}\sigma_{N}}\exp\left(-\frac{x^{2}}{2\sigma_{N}^{2}}\right)\\
\epsilon\left(x\right) & = & \epsilon_{0}\exp\left(-\frac{x^{2}}{2\sigma_{\epsilon}^{2}}\right)\\
C\left(x_1,x_2\right) & = & C_{0}\exp\left(-\frac{\left(x_1-x_2\right)^{2}}{2\sigma_{c}^{2}}\right),
\end{eqnarray}
and that the bins are sufficiently small so that we may replace the sums by integrals.

Typically the range of the correlation functions is expected to be shorter than that of the particles distribution, $r_c\equiv \sigma_c/\sigma_N <1$. On the other hand $r_\epsilon \equiv \sigma_\epsilon/\sigma_N$ depends on the specific detector system. In addition to the various Gaussians, the overall acceptance of the detector needs to be taken into account as well, and we denote the interval in phase space where particles are actually measured by $x \in (-\Delta,\Delta)$.  This will determine the range of integration (summation) for the above expressions. If the acceptance $\Delta$ is comparable to the range of the particle distribution we may integrate over the full phase-space, $x \in (-\infty,\infty)$ and we get
\begin{equation}
R\equiv\frac{\overline{\delta}}{\delta}=\sqrt{\frac{\left(2+r_{c}^{2}\right)}{r_{c}^{2}+\frac{2r_{\epsilon}^{2}}{\left(1+r_{\epsilon}^{2}\right)}}}.
\label{eq:ratio_1}
\end{equation}
If the efficiency changes only little over the range of the particle distribution than the effect of local efficiency corrections is negligible, and in the limit of $r_{\epsilon }\rightarrow \infty $ we recover the result for constant efficiency, i.e. $R=1$. If, on the other hand, the range of the efficiency corrections are smaller than or comparable with that of the particle distribution, $r_\epsilon \lessapprox 1$, the correction become significant, $R\gtrapprox \sqrt{2}$. However, in this is case a more quantitative estimate require to account for the overall acceptance, $\Delta$, since in this limit we will have particles in the region with vanishingly small efficiency, i.e. no acceptance. 

If the measurement is performed in the region smaller than the range of the particle distribution we may only integrate over $x \in ( -\Delta,\Delta)$. In this case no simple analytical expression can be obtained for $R$. Numerical studies show that for example $R=\bar{\delta}/\delta \approx 1.25$ if we assume that the efficiency is 20\% at the boundary of the acceptance, $\epsilon(\Delta) = 0.2$. In other words, global efficiency correction leads to 25\% larger deviations from the Poisson limit than the local efficiency correction.  Although, one would expect the effect to increase for higher order cumulants, it is difficult to imagine factors of 2 or more due to the neglect of local efficiency corrections. We note, that the signal for a potential phase structure is the deviation from Poisson behavior and, therefore, the above corrections, while not tremendous, are still significant and need to be properly accounted for. 

\section{Concluding Remarks}
\label{sec:concluding-remarks}
Let us conclude with a few remarks. 
\begin{enumerate}
\item We note that the above result, Eq.~\eqref{eq:final_result}, does not require the phase-space
  bins to be of equal size. Thus Eq.~\eqref{eq:final_result} applies
  to any choice of binning most suitable for a given experiment.
\item In the limit of constant efficiency over the entire phase-space
  under consideration, our result, Eq.~\eqref{eq:final_result},
  reduces to  Eq.~\eqref{eq:6}, the consequences of which were subject
  of our previous paper \cite{Bzdak:2012ab}.
\item As already pointed out at the beginning of the paper, the above
  results assume a binomial distribution as a model for particle
  detection efficiencies. Thus the above expressions need to be
  suitably modified if this is not the case in a given experiment. 
\end{enumerate}

\section*{Acknowledgments} 
%%%%%%%%%%%%%%%%%%
\hspace*{\parindent}
A.B. was supported through the RIKEN-BNL Research Center and Grant No. UMO-2013/09/B/ST2/00497. V.K. was supported by the Office of Nuclear Physics in the US Department of Energy's Office of Science under Contract No. DE-AC02-05CH11231.

\end{document}